\documentclass[dvips]{article}
\usepackage{icrctc07}

\newcommand{\beq}{\begin{equation}}
\newcommand{\eeq}{\end{equation}}
\newcommand{\bea}{\begin{eqnarray}}
\newcommand{\eea}{\end{eqnarray}}
\newcommand{\bweq}{\begin{widetext}\begin{equation}}
\newcommand{\eweq}{\end{equation}\end{widetext}}
\newcommand{\bwt}{\begin{widetext}}
\newcommand{\ewt}{\end{widetext}}
\newcommand{\ovl}{\overline}

\newcommand{\MeV}{\mbox{\ MeV}}
\newcommand{\GeV}{\mbox{\ GeV}}
\newcommand{\TeV}{\mbox{\ TeV}}
\newcommand{\PeV}{\mbox{\ PeV}}
\newcommand{\EeV}{\mbox{\ EeV}}

\newcommand{\mum}{\mbox{\ $\mu$m}}
\newcommand{\mb}{\mbox{\ mb}}

\title{Prompt muons in extended air showers}

\shorttitle{Prompt muons}

\authors{J. Ridky$^{1}$, D. Nosek$^{2}$, P. Travnicek$^{1}$, P. Necesal$^{1}$} 

\shortauthors{D. Nosek and et al}

\afiliations{
$^{1}$ Institute of Physics of the Academy of Sciences of the Czech Republic, Prague, Czech Republic\\
$^{2}$ Institute of Particle and Nuclear Physics, Charles University, Prague, Czech Republic
}

\email{ridky@fzu.cz, nosek@ipnp.troja.mff.cuni.cz}

\abstract{We present results of simulations of a muon content in the air
showers induced by very high energy cosmic rays. 
Muon energy distributions and muon densities at ground level are given. 
We discuss a prompt muon component generated by decays of charm mesons. 
The method combines standard Monte Carlo generators incorporated in 
the CORSIKA code and phenomenological estimates of the charm hadroproduction.}

\begin{document}
\maketitle


\section{Introduction}

Extended air showers (EAS) are initiated by very energetic primary 
cosmic ray (CR) particles interacting with air by producing many
secondary hadrons that may produce other hadrons in subsequent 
interactions.
In the EAS development a special role is played by neutral pions that
rapidly decay into gammas.
These gammas start an electromagnetic shower carrying typically 90\%
of the initial energy.
The rest of the energy showed up as muons from hadronic decays.
Whereas the electromagnetic part of the shower is well understood, 
the muon component depends strongly on the mechanism of the hadronic 
interactions, properties of which are not well known at incident 
energies above a few hundred of~\GeV~in c.m.s.

Available data in the~\GeV--\TeV~energy range obtained with surface 
and underground detectors are still too discrepant to draw definite
conclusions on the muon spectra induced in EAS.
It is known that direct measurements of low energy
muons~\cite{Ant01,Abu01} as well as experimental data on high 
energy muons~\cite{Ava01,Rid01,L3C01} 
are not described satisfactorily by 
simulations using the currently employed hadronic interactions 
model~\cite{Kal1,Ost01,Eng01,Ran01}.
Some interesting features about the EAS muon production have been
recently obtained using the high energy hadronic interaction model 
EPOS~\cite{Pie01}.
A preliminary analysis of EPOS results showed that due to the enhanced 
(anti)baryon production the number of muons in the EAS increases more rapidly 
with energy than in the currently used high energy hadroproduction models.
Nonetheless, one cannot exclude the possibility that something important 
is missing in interaction models, especially, concerning the very high 
energy region.

In this work, we briefly discuss the relationship between the hadronic
multiparticle production and EAS observables.
A special attention is paid to charm secondaries that are accessible in 
hadronic interactions during the EAS development initiated by very high
energy CR primaries, typically at energies greater than 1\PeV~\cite{Nec01}.
The main goal is to discuss the impact of the production of charm
particles and their prompt decays on the EAS muon content that is
relatively easily measured with great accuracy.  


\section{Model of prompt muon production}

The number of muons registered by a ground array is one of the most
important observables in EAS physics.
It depends on the primary energy and the details of consecutive 
collisions of shower hadrons with air nuclei.

At~\GeV~energies the EAS muon component is dominated by conventional
sources, i.e. the weak decays of relatively long--lived mesons, pions 
and kaons.
At very high energy pion or kaon decays become very rare.
For energies of the order of 1\TeV~and greater, the probability increases 
that such particles interact in the atmosphere before decaying.
This implies that even a small fraction of short--lived mesons and
baryons containing heavy quarks, most notably charm, decaying into muons 
can give the important contribution to high energy EAS muons.

All the high--energy interaction models reproduce reasonably well
accelerator data but differ in their predictions above few~\TeV~in c.m.s.
Specifically, the charm production is strongly suppressed in hadronisation 
models that are commonly used in EAS physics.
A detailed critical discussion of the best possible choice of the
charm production model and of the systematic uncertainties connected
with it is beyond the scope of this study, for more details 
see e.g.~\cite{Nov01}.
Here, we use a simple phenomenological model of the charm hadroproduction. 

We have modified the mechanism of hadronic collisions artificially 
to gain a number of charm particles.
It is assumed that high energy hadrons that are present in the EAS core
interact actively when they pass through the atmosphere and aside
an abundant pion and kaon production also charm particles are produced 
in some events.
The charm cross section was adjusted in accord with recent 
measurements~\cite{Zho01,Kel01} that for the nucleon--nucleon 
collision yield $\sigma_{c\ovl{c}} \approx 1\mb$ 
at energies of 200\GeV~in c.m.s.
We assume that the charm cross section in p-Air
collisions grows up logarithmically~\cite{Nov01}, 
$\sigma_{c\ovl{c}} \approx 0.02 \ \sigma_{{\rm inel}}$ 
at the incident energy of 100\TeV~reaching a value 
$\sigma_{c\ovl{c}} \approx 0.06 \ \sigma_{{\rm inel}}$ 
slightly above the incident energy of 10\EeV, see also~\cite{Dre01}. 

In the present approximation, mesons and nucleons are generated 
in nucleon--air collisions and in consecutive hadron--air collisions.
Part of the initial energy of the interaction is carried by the involved 
conventional hadrons ($p, n, \pi, K, \eta$).
The number of these hadrons remains unchanged during the collision, their energies
are degraded and a modest fraction of their total energy, typically
less than 20\%, is transformed into the energy of final charm degrees 
of freedom.
Energy spectra of produced charm particles are mimicked using spectra 
of ordinary secondaries.
Because at the energy of interest their secondary interactions seldom occur, 
it is assumed that charm particles ($D, \Lambda_{\rm c}$) decay very rapidly
($c\tau \approx 50-300 \mum$).
Even though there are many semi--leptonic decay channels for charm particles 
and most of them have more than three particles in the final state we do not
investigate these processes in detail.
We assume that charm particles decay mostly into muons with typical
branching ratios for semi--leptonic decays 
$\rm{BR}(D, \Lambda_{c} \to e,\mu) \approx 10-20\%$; 
also hadronic decay modes producing secondary mesons are included in the model.
\begin{figure}
\begin{center}
\includegraphics [width=0.48\textwidth]{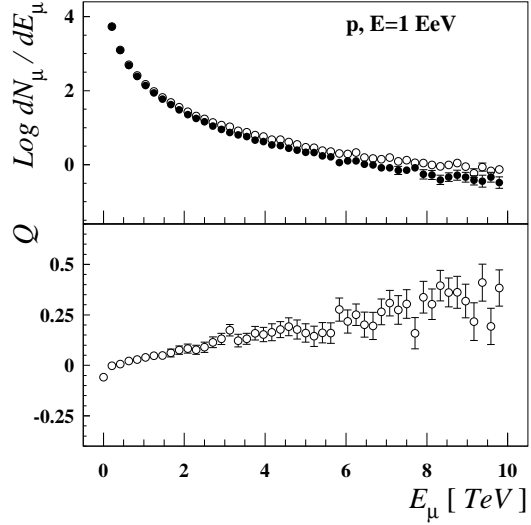}
\end{center}
\vspace{-0.5cm}
\caption{Ground level energy spectra of muons originated in EAS with 
  (open circles) and without (full circles) charm production are shown 
  in upper panel. 
  The incident proton energy is set to 1\EeV. 
  Corresponding excess of muons born in EAS with charm production is shown 
  in lower panel as a function of muon energy.
}
\label{F01}
\end{figure}


\section{EAS simulations}

Being formed during a multistep hadronic cascade, the EAS muon content
is closely connected to the mechanism of hadron--air collisions.
These collisions are investigated in the standard treatment.
To obtain the energy spectra of conventional particles the QGSJET01 
model~\cite{Kal1} of hadronic interactions is employed.
The GHEISHA procedure is used to treat hadronic collisions of secondary 
particles at small energies.
To describe the propagation of the particles in EAS down through 
the atmosphere the EAS simulation code CORSIKA~\cite{Hec1} is adapted.
\begin{figure}
\begin{center}
\includegraphics [width=0.43\textwidth]{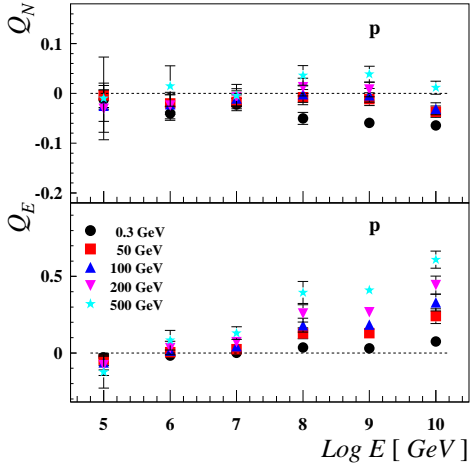}
\end{center}
\vspace{-0.5cm}
\caption{Ground level excesses of the total number (upper panel) and
  energy (lower panel) of muons in EAS with charm production are
  depicted as functions of the primary proton energy. 
  Muon energies are constrained as $E_{\mu} \ge 0.3, 50, 100, 200$ and $500 \GeV$.
}
\label{F02}
\end{figure}
\begin{figure}
\begin{center}
\includegraphics [width=0.43\textwidth]{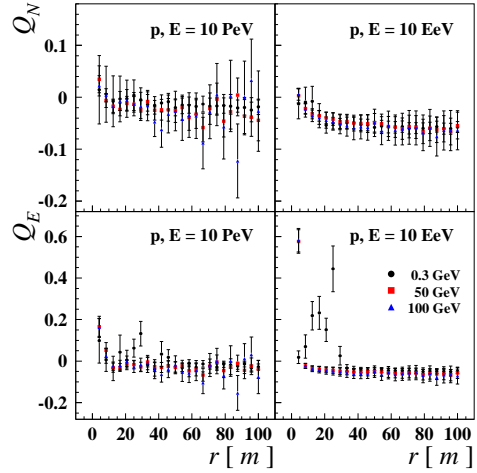}
\end{center}
\vspace{-0.5cm}
\caption{Ground level lateral excesses of the number (upper panels)
  and energy (lower panels) of muons in EAS with charm production
  initiated by the proton primary with the incident energy of 
  10\PeV~and 10\EeV~are depicted.
  Results for muon energies $E_{\mu} \ge 0.3, 50$ and $100 \GeV$ are shown.
}
\label{F03}
\end{figure}

Charm particles are produced at the expense of the energy of
secondary baryons and mesons produced in subsequent hadron--air 
collisions in which the interaction energy exceeds 150\GeV~in c.m.s.; 
no account of their energy spectra is taken.
Electron--photon cascades due to the decays of neutral pions are 
treated by common standards.

The CR primary protons are assumed to interact with the air nuclei 
at various incident energies of 100\TeV--10\EeV.
The primary zenith angle is fixed at zero degrees; the altitude of 
the initial interaction is left free.
In all simulations, a standard U.S. atmosphere is used.

The kinetic energy cutoffs for EAS hadrons and muons are chosen 
to 0.3\GeV; for electrons/positrons and photons we use cutoffs 
of 20\MeV~and 2\MeV, respectively.
In all calculations a thinning level of $10^{-7}$ and a maximum weight
factor of $10^{30}$ are adopted for both electromagnetic as well
as hadronic particles.
In order to minimise influence of shower to shower fluctuations 
we average over 100 air showers; statistical uncertainties are
shown in figures.

To visualise effects associated with the EAS charm production 
we use the asymmetry--like quantity
$Q = \frac{N_{\mu}' - N_{\mu}}{N_{\mu}' + N_{\mu}}$,
where $N_{\mu}'$ and $N_{\mu}$ are the number or energy densities of muons 
originating respectively in EAS with or without charm production and
registered by the ground detector.
This quantity measures an excess of muons originated in EAS with charm
production over muons in showers generated in conventional models.


\section{Numerical results}

Depending on the primary energy, a fraction of charm particles with
respect to hadrons produced in all consecutive collisions during 
the EAS development is $10^{-5} - 10^{-4}$ in our calculations.
An energy fraction carried by these charm particles increases more 
rapidly with the increasing primary energy from a value of $10^{-5}$ 
at the~\PeV~region reaching $10^{-2}$ at the energy of 10\EeV.

The energy spectra of muons initiated by the protons of the primary
energy of 1\EeV~and detected at the ground as obtained 
in our simulations with the QGSJET01 high--energy interaction model 
are depicted in Fig.\ref{F01}.
Depending on the initial energy, the energy spectra of muons
fall off by about 4--6 orders of magnitude in the 10\GeV--10\TeV~range. 
These spectra have typical profiles with a visible excess of 
hard muons due to charm production.
At the primary energies of interest, secondary pions and kaons 
are mostly above the critical energy ($E_{\rm cr} < 1\TeV$) and so 
predominantly generate conventional muons in interactions, 
while charm particles being below their critical energy 
($E_{\rm cr} > 10\PeV$) give prompt muons of high energies. 
This effect is remarkably large reaching a factor of two for highest 
muon energies studied.

Some general features of our simulations that can be observed by 
the ground level detector or underground are summarised 
in Fig.\ref{F02}.
Here excesses of the total number and energy of muons originated 
in EAS with charm production are shown as functions of the incident 
energy of the CR proton. 
The total number of muons and their energy are depicted for muons with
energies $E_{\mu} \ge 0.3, 50, 100, 200$ and $500 \GeV$ 
that in our calculations imitate the rock overburden for underground 
experiments.

It is well visible that muons in EAS with charm production deposit
remarkably more energy in the detector than muons in conventional showers.
This effect is pronounced with the increasing lower energy cut for muons.
Our simulations show that charm particles carrying away 
a non--negligible fraction of the incident energy can be 
responsible for the delay of the energy absorption along 
the EAS axis.
On the other hand, the number of muons born in EAS with charm production 
and registered at the ground or underground remains relatively stable 
with the increasing primary energy and, although mostly smaller, 
it does not differ considerably from the number of muons in showers 
generated in conventional models. 

Examples of the lateral number and energy densities of muons initiated 
by the protons of the primary energies of 10\PeV~and 10\EeV~are 
depicted in Fig.\ref{F03}.
We have found a small but visible deficit of muons in very high
energy EAS with charm production with respect to conventional showers.
On contrary, due to charm production an excess of muon energy 
that should be deposited near the shower axis in the ground detector 
or underground is observed.


\section{Conclusions}

The calculation of the EAS muon content suffers in principle a
significant uncertainty due to the lack of knowledge of the properties 
of the charm production in the hadron--nucleus collisions.
We showed that a simple phenomenological treatment of
prompt muons can reveal interesting observable features.
More precise analysis will be carried out in the future.


\section{Acknowledgements}
This work is supported by the Ministry of Education, Youth and Sports
of the Czech Republic under Contracts Nos.~LC~527 and MSM~0021620859.




\end{document}